\documentclass[twocolumn,epjc3,final]{svjour3}  
 
\usepackage{amsmath,amssymb,amsfonts,dsfont, braket}             
\usepackage{graphics}
\usepackage{graphicx}
\usepackage{lineno}
\usepackage{hyperref}

\usepackage{xcolor}
\newcommand{\comments}[1]{{#1}}
\newcommand{\Lorenzo}[1]{{#1}}
\newcommand{\New}[1]{{#1}}

\begin{document}


\title{Geometrical Constraints On Leptonic Unitarity Triangles}


\author{Mathieu Guigue \thanksref{e1,addr1} 
\and
Lorenzo Restrepo \thanksref{e2,addr1,addr2} 
}
\thankstext{e1}{e-mail: mathieu.guigue@lpnhe.in2p3.fr}
\thankstext{e2}{e-mail: lorenzo.restrepo@lpnhe.in2p3.fr}


\institute{Laboratoire de Physique Nucléaire et des Hautes Énergies, Sorbonne Université, Paris, France \label{addr1}
           \and
           IPSA-DRII,  63 boulevard de Brandebourg, 94200 Ivry-sur-Seine, France \label{addr2}
}

\date{\today}
\maketitle
\begin{abstract}
The precision of the neutrino oscillation parameters measurements has improved and will continue to improve as the next-generation experiments become online.
Beyond the more precise measurements of the mixing angles and phases used to parametrize the lepton mixing matrix, tests of its unitarity are of great interest.
This paper studies how the amplitudes of the oscillation patterns can be used and combined to construct leptonic unitarity triangles.
\end{abstract}

\section{\label{sec:unitarity}Unitarity triangles in the lepton sector}

\comments{The oscillation of the neutrinos from one flavour to another has been a well-established quantum phenomenon for the last three decades, since its discovery by the SNO \cite{Ahmad2002} and Super-Kamiokande \cite{Fukuda1998} experiments.
These oscillations are generally described within the Pontecorvo-Maki-Nakagawa-Sakata (PMNS) framework\Lorenzo{, \cite{Maki1962,Gribov1969} which involves a \comments{$3\times 3$} unitary mixing matrix, for the three known neutrino types.}
}

As neutrino oscillation physics enters a precision era, it can be of interest to test our more fundamental hypotheses, such as the unitarity of the PMNS matrix.
The presence of heavier neutrinos could be described by a larger dimension mixing matrix from which only a $3\times 3$ subset is the PMNS matrix: \comments{typically, the see-saw models \cite{Fritzsch1975} that would explain the neutrinos mass origin \Lorenzo{in a natural  way,} predict the existence of several heavy neutrinos mixing with the standard light neutrinos \cite{Antusch2006}.}
In such case, one can imagine that the $3\times 3$ matrix describing the standard neutrinos oscillation is not unitarity, and \Lorenzo{thus} the equation
\begin{eqnarray}\label{eq:unitarity-property}
	UU^\dagger =U^\dagger U = \mathds{1}_3,
\end{eqnarray}
doesn't hold.
There are various ways to test unitarity, \comments{in particular via the search of suppressed leptons decays e.g. $\mu \rightarrow e\gamma$.
The resulting constraint consists in the direct constraint on the off-axis terms of $U^\dagger U$ \cite{Antusch2006}.
Alternatively,} one can constrain the mixing matrix element ratios on the complex plane: these are called ``unitarity triangles".
Such search is done extensively in the quark sector, where the multiple probes that exist allow to tightly constraint the allowed region for the apexes of these triangles \cite{CKMFitter2005}.

This paper presents how constraints on unitarity triangles in the lepton sector can be obtained from the individual neutrino oscillation measurements without assuming unitarity of the PMNS matrix.
Section \ref{sec:probability} presents the oscillation probability expression useful for this work. 
\comments{Section \ref{sec:construction} describes the general procedure for constructing unitarity triangles.
The connection between the neutrino oscillation amplitudes and the unitary triangle is described in Section \ref{sec:geometry}, while the constructed triangles are displayed in Section \ref{sec:implementation-stan}.
Finally, Section \ref{sec:discussion} discusses the difficulty of extracting the oscillation amplitudes and therefore the construction of unitarity triangles, while providing some insights about possible remedies.}

\section{\label{sec:probability}Neutrino oscillation and oscillation amplitudes}

\subsection{Neutrino oscillation probability}

\Lorenzo{Neutrino oscillation arises from a mismatch between the neutrino flavour eigenstates $\ket{\nu_{\alpha}}$ by which neutrinos interact, and mass eigenstates $\ket{\nu_i}$ by which they propagate. Flavour eigenstates can be decomposed as a linear combination of mass eigenstates:}
\begin{eqnarray}
\ket{\nu_{\alpha}} = \sum_{i = 1}^{N} U_{\alpha i}^{*} \ket{\nu_{i}} ,
\end{eqnarray}
\noindent with $N \geq 3$ being the number of neutrino mass eigenstates and $U$ being the generic lepton mixing matrix.

\Lorenzo{The} probability of a neutrino \Lorenzo{of} flavour \Lorenzo{$\alpha$} to oscillate into \Lorenzo{the} flavour \Lorenzo{$\beta$} depends \Lorenzo{exclusively} on the mixing matrix elements, on the neutrino energy $E$ and on the propagation length $L$. \Lorenzo{In all generality neutrino oscillation probability in vacuum can be expressed as: \begin{eqnarray}\label{eq:general_oscprob}
P_{\alpha\beta} &=& \frac{1}{n_{\alpha}n_{\beta}} [\: |\sum_{j}U_{\alpha j}^{*}U_{\beta j}|^2 
\nonumber\\
&-& \sum_{j < k} 4 \, \mathcal{R}e[U_{\alpha j}^{*} U_{\beta j} U_{\alpha k} U_{\beta k}^{*}] \sin^2(\frac{\Delta m_{j k}^2 L}{4 E})
\nonumber\\ &+& \sum_{j < k} 2 \, \mathcal{I}m[U_{\alpha j}^{*} U_{\beta j} U_{\alpha k} U_{\beta k}^{*}]\sin(\frac{\Delta m_{j k}^2 L}{2 E}) \; ], \end{eqnarray}}
\noindent\Lorenzo{with the normalisation factor $n_{\alpha} = |\braket{\nu_{\alpha}|\nu_{\alpha}}|$. If the sum covers indeed all the possible neutrino families (including sterile neutrinos), then the normalisation factor should be equal to one $n_{\alpha} = 1$. Nevertheless, some authors prefer to leave the normalisation free to be as conservative as possible \cite{Ellis2020b}. On the other hand, the antineutrino oscillation probability is expressed like Eq.~\eqref{eq:general_oscprob} but applying the complex conjugation to every matrix element $U_{\alpha i}$.}\\ 

Starting from Eq.~\eqref{eq:general_oscprob}, there can be various ways to test the unitarity of the $3 \times 3$ PMNS matrix $U$. \comments{For example, one can} incorporate one or more sterile neutrinos in the oscillation probability and use this model within the oscillation analysis frameworks \cite{Denton2021,Coloma_2021}. \New{One of the most widely used approaches in the literature is the Minimal Unitarity Violation scheme (MUV) defined in reference \cite{Antusch_2006}. It consists on considering only three light neutrino species that mix via a potentially non-unitary mixing matrix. Some authors choose to parametrize these deviations from unitarity of $U$ and try to constrain these new parameters within the oscillation analysis frameworks \cite{Trzeciak_2025}. In this paper we also choose to work under the MUV scheme, considering only three active neutrino species, but without parametrizing the unitarity deviations. We will therefore only focus on the neutrino oscillation amplitudes \cite{Parke_2016} and the mixing matrix elements themselves.}

\subsection{Oscillation probability amplitudes}

\New{The three-neutrino oscillation probability \eqref{eq:general_oscprob} can be written in a compact way once we introduce the oscillation amplitudes. For the neutrino disappearance we define the amplitudes $A_{\alpha, ij}$ as:
\begin{eqnarray}
A_{\alpha,ij} = \vert U_{\alpha i}\vert ^2\vert U_{\alpha j}\vert ^2
\end{eqnarray}
with $\alpha \in\left\{ e,\mu,\tau\right\}$, in such a way that we can rewrite the neutrino disappearance probability as:
\begin{eqnarray}
P_{\alpha \alpha} =  |\sum_{j}U_{\alpha j}^{*}U_{\beta j}|^2 
- 4 \,\sum_{j < k} \, A_{\alpha, j k} \, \sin^2\left(\frac{\Delta m_{j k}^2 L}{4 E}\right) .
\end{eqnarray}
}

\New{The first term of the probability is a normalization factor that is equal to 1 in the case of unitary matrix as of Eq.~\eqref{eq:unitarity-property}.}

\New{Similarly, for the neutrino appearance we define the quantities $\mathcal{R}_{\alpha\beta,ij}$ and $\mathcal{I}_{\alpha\beta,ij}$ as:
\begin{eqnarray}
	\mathcal{R}_{\alpha\beta,ij} = \mathcal{R}e\left(U^*_{\alpha i}U_{\beta i} U_{\alpha j} U^* _{\beta j} \right)\\
	 \mathcal{I}_{\alpha\beta,ij} = \mathcal{I}m\left(U^*_{\alpha i}U_{\beta i} U_{\alpha j} U^* _{\beta j} \right).
\end{eqnarray}
}

\New{With this notation the neutrino appearance probability can be rewritten as:
\begin{eqnarray}
P_{\alpha\beta} = &-&4 \, \sum_{j < k} \, \mathcal{R}_{\alpha\beta,jk} \, \sin^2\left(\frac{\Delta m_{j k}^2 L}{4 E}\right)  \, \nonumber \\ &+& 2\, \sum_{j < k} \, \mathcal{I}_{\alpha\beta,jk} \, \sin\left(\frac{\Delta m_{j k}^2 L}{2 E}\right).
\end{eqnarray}
Note how $\mathcal{R}_{\alpha\beta,ij}$ is symmetric in $\alpha$ and $\beta$ as well as $i$ and $j$, and how $\mathcal{I}_{\alpha\beta,ij}$ is antisymmetric in the same indexes.
}

\New{Let us apply this notation for the first two rows of the mixing matrix.} The \Lorenzo{three} neutrino oscillation probabilities can be made explicit for muon neutrino disappearance ($\nu_{\mu}\rightarrow\nu_{\mu}$), electron neutrino disappearance ($\nu_{e}\rightarrow\nu_{e}$) and electron neutrino appearance ($\nu_{\mu}\rightarrow\nu_{e}$):
\New{\begin{eqnarray}
	P_{\mu\mu} =& & \left(\sum_{k}\vert U_{\mu k}\vert^2\right)^2  - 4 \, A_{\mu, 12}  \sin^2\left(X_{12}\right) \nonumber\\&-& 4 \, A_{\mu, 13} \sin^2\left(X_{13}\right) - 4 \, A_{\mu, 23} \sin^2\left(X_{23}\right), \label{eq:P_mu_mu}
\end{eqnarray}
\begin{eqnarray}
	P_{ee} = & & \left(\sum_{k}\vert U_{e k}\vert^2\right)^2 - 4  \, A_{e, 12} \sin^2\left(X_{12}\right)
	\nonumber\\&-& 4  \, A_{e, 13}  \sin^2\left(X_{13}\right) 
	- 4  \, A_{e, 23} \sin^2\left(X_{23}\right),\label{eq:P_e_e}
\end{eqnarray}\\
\begin{eqnarray}
	P_{\mu e} =  &-& 4 \, \mathcal{R}_{\mu e,12}  \sin^2\left(X_{12}\right)+ 2 \, \mathcal{I}_{\mu e,12} \sin\left(2X_{12}\right)
	\nonumber\\&-& 4 \, \mathcal{R}_{\mu e,13} \sin^2\left(X_{13}\right)\nonumber + 2 \, \mathcal{I}_{\mu e,13} \sin\left(2X_{13}\right)
	\nonumber\\&-& 4 \, \mathcal{R}_{\mu e,23}\sin^2\left(X_{23}\right)+ 2 \mathcal{I}_{\mu e,23} \sin\left(2X_{23}\right), \label{eq:P_mu_e}
\end{eqnarray}}
with $X_{ij} = \frac{\Delta m^{2}_{i j} L}{4 E}$ \comments{and $\Delta m^{2}_{i j} = m^2_i-m^2_j$ being the squared mass difference between two mass eigenstates $i$ and $j$}.

Experimentally, if one is capable of measuring these oscillation probabilities as a function of $X_{ij}$, the amplitudes of the oscillations could be extracted.
\comments{For example, in the case of the muon neutrino disappearance, one could extract independent constraints on $A_{\mu, 12}$, $A_{\mu, 13}$, and $A_{\mu, 23}$.} In reality, as it was pointed out in Ref. \cite{Ellis2020}, given the values of the squared masses splitting, it might be difficult to disentangle between the $\sin^2\left(X_{13}\right)$ and $\sin^2\left(X_{23}\right)$ terms, and therefore measure independently their respective amplitudes. In the interest of the discussion, we will assume one can distinguish between these \New{while leaving further discussions on this particular point to Section \ref{sec:discussion}}.

\section{Relationship between oscillation probability amplitudes and unitarity triangles}

\subsection{\label{sec:construction}Unitarity triangles construction}

\comments{By construction,} a $3\times 3$ complex matrix has 18 independent parameters.
\comments{In particle physics and in the context of the lepton mixing,} additional constraints on these parameters can be added, such as the removal of unphysical phases due to lepton field rephasing.
For now, we consider that the non-unitary neutrino mixing matrix has indeed 18 independent parameters.
Testing the unitarity \eqref{eq:unitarity-property} of a matrix consists then in testing the validity of the following equations:
\begin{eqnarray}
	\sum _{\alpha} \vert U_{\alpha i}\vert ^2 &=& 1,\label{eq:unit-norm-row}\\
	\sum _{i} \vert U_{\alpha i}\vert ^2 &=& 1,\label{eq:unit-norm-column}\\
	\sum _i U_{\alpha i}U^*_{\beta i} &=& 0 ~(\alpha\neq\beta),\label{eq:unit-closure-row}\\
	\sum _\alpha U_{\alpha i}U^*_{\alpha j} &=& 0~(i\neq j).\label{eq:unit-closure-column}
\end{eqnarray}
Equations \eqref{eq:unit-norm-row} and \eqref{eq:unit-norm-column} relate to the probability normalization \cite{Ellis2020b}, while Eq. \eqref{eq:unit-closure-row} and \eqref{eq:unit-closure-column} are used to produce the so-called \textit{leptonic unitarity triangles} as these equations can be seen from a geometrical point of view in the complex plane.
In the case of 3 neutrino flavours and 3 mass eigenstates, these conditions can produce up to 6 triangles.
Let's consider the following condition as one possible case:
\begin{eqnarray}\label{eq:unit-condition-emu}
	\frac{U_{e1}U^*_{\mu 1}}{U_{e3}U^*_{\mu 3}} +\frac{U_{e2}U^*_{\mu 2}}{U_{e3}U^*_{\mu 3}} + 1 = 0 
\end{eqnarray}
which corresponds to Eq. \eqref{eq:unit-closure-row} for $\alpha=e$ and $\beta=\mu$.
Usually, one defines the quantities\footnote{For convenience, we usually drop the $^{(1)}$.} $\rho^{(1)}_{e\mu}$ and $\eta^{(1)}_{e\mu}$ as, respectively, the negative value of real and imaginary parts of the first ratio in Eq. \eqref{eq:unit-condition-emu}: 
\begin{eqnarray}\label{eq:rho_eta_def}
	\rho_{e\mu}+i\eta_{e\mu} = -\frac{U_{e1}U^*_{\mu 1}}{U_{e3}U^*_{\mu 3}}.
\end{eqnarray}
The coordinates of $\rho_{e\mu} +i\eta_{e\mu}$ in the complex plane corresponds to the apex of a triangle, the other two being (0,0) and (0,1), one of the sides of the triangle being normalized to a length of 1 by convention.
\Lorenzo{Note that the apex of the triangle can also provide information about CP violation. Indeed, if the apex is not on the real axis, then there must be an imaginary component coming from at least one CP violating parameter of $U$.} We can also define the quantities $\rho'_{e\mu}$ and $\eta'_{e\mu}$ as:
\begin{eqnarray}\label{eq:rho_eta_prime_def}
	\rho'_{e\mu}+i\eta'_{e\mu} = \frac{U_{e2}U^*_{\mu 2}}{U_{e3}U^*_{\mu 3}} + 1.
\end{eqnarray}
Testing the unitarity for this triangle therefore consists in verifying that the apex defined by $\rho_{e\mu}+i\eta_{e\mu}$ superimposes with the one defined by $\rho'_{e\mu}+i\eta'_{e\mu}$.
\comments{Any deviation from unitarity would graphically cause a non-superimposition of the two apexes.}
\Lorenzo{The former} triangle is \Lorenzo{particularly} interesting as it involves only the electron and muon neutrinos\Lorenzo{, to which }long-baseline experiments are sensitive \Lorenzo{and could therefore constrain. }


\subsection{\New{Unitarity studies in the literature and the need for experimental constraints}}

Neutrino oscillation experiments measure the oscillation probability usually as a function of the neutrino energy or the propagation length; therefore they are sensitive to the oscillation amplitudes $A_{\alpha,ij}$ in the case of disappearance experiments or $\mathcal{R}_{\alpha\beta,ij}$ and $\mathcal{I}_{\alpha\beta,ij}$ in the case of appearance experiments.
As these experiments are mostly dedicated to the measurement of the mixing angles within the PMNS framework, they don't report their measurements of the oscillation amplitudes.
Unitarity tests using these measurements have been extensively studied in the literature, especially in the context of the future long-baseline neutrino experiments \cite{Farzan2002,Ellis2020b}, 
\New{using various analysis strategies. 
Most of the authors choose to interpret the reported neutrino spectra from the experiments directly in terms of oscillation amplitudes that are in turn interpreted as unitarity constraints \cite{Parke_2016,Kitano2026,Gonzalez-Garcia2014}. Some authors employ this type of approach, further solidifying their analysis by using priors on some of the mixing angles and mass squared splittings \cite{Ellis2020,Ellis2020b}. An important caveat of most analyses is the approximations that are done, such as not taking into account the individual systematics for each experiment, or sometimes assuming fixed-energy neutrino beams.}

\New{Instead, it would be preferable for the measurements of the oscillation amplitudes to be directly reported by the collaborations, as the result of a thorough analysis taking into account all the systematic effects and specificities of each experiment. In the precision era of neutrino physics, the experimental involvement in measuring the combinations of oscillation amplitudes will be crucial to test the unitarity of the lepton mixing matrix, as shown in the next sections.}

\subsection{\label{sec:geometry}Geometrical constrains on the unitarity triangle from oscillation amplitudes}

\comments{A natural question is therefore how the constraints on the oscillation amplitudes will appear in a $(\rho_{e\mu},\eta_{e\mu})$ plane, similarly to the CKM one \cite{CKMFitter2005}}.
For example, what combination of constraints is required to produce a circle on this plane? 
Let's compute the radius $R$ of a circle centred on $(\rho_{e\mu},\eta_{e\mu})=(0,0)$ and see how it can be expressed as a function of the oscillation amplitude constraints.
The radius $R$ is defined as
\begin{eqnarray}
	R^2 &=& \rho_{e\mu}^2+\eta_{e\mu}^2
    = \frac{\vert U_{\mu 1}\vert ^2\vert U_{e 1}\vert ^2}{\vert U_{\mu 3}\vert ^2\vert U_{e 3}\vert ^2},
\end{eqnarray}
which can be written as 
\begin{eqnarray}\label{eq:unit_radius}
	R^2 = \rho_{e\mu}^2+\eta_{e\mu}^2 = \frac{\vert U_{\mu 1}\vert ^2\vert U_{\mu 2}\vert ^2\vert U_{e 1}\vert ^2\vert U_{e 2}\vert ^2}{\vert U_{\mu 2}\vert ^2\vert U_{\mu 3}\vert ^2\vert U_{e 2}\vert ^2\vert U_{e 3}\vert ^2} = \frac{A_{\mu,12}A_{e,12}}{A_{\mu,23}A_{e,23}}.
\end{eqnarray}
This expression shows that the constraints on several disappearance amplitudes can be combined into a circle in the $(\rho_{e\mu},\eta_{e\mu})$ plane.
Similarly, one can combine the constraints $A_{\mu,12}$, $A_{e,12}$, $A_{\mu,13}$ and $A_{e,13}$ into a circle centered on $(\rho'_{e\mu},\eta'_{e\mu})=(1,0)$ with a radius $R'$:
\begin{eqnarray}\label{eq:unit_radius_prime}
	R'^2 &=& (\rho'_{e\mu}-1)^2+\eta'^2_{e\mu} \nonumber \\
    &=& \frac{\vert U_{\mu 1}\vert ^2\vert U_{\mu 2}\vert ^2\vert U_{e 1}\vert ^2\vert U_{e 2}\vert ^2}{\vert U_{\mu 1}\vert ^2\vert U_{\mu 3}\vert ^2\vert U_{e 1}\vert ^2\vert U_{e 3}\vert ^2} = \frac{A_{\mu,12}A_{e,12}}{A_{\mu,13}A_{e,13}}.
\end{eqnarray}
From these expressions of $R$ and $R'$, in order to constrain each circle one needs to measure 4 disappearance amplitudes.
Moreover, it shows that the uncertainties on the \Lorenzo{circle} radius are directly related to the precision on each amplitude measurement; in other words, the precision on the radius is driven by the least precise amplitude measurement.

Constraints on the angles of the triangle or similarly on the slope of its sides are also possible as they arise from a constraint of $\eta_{e\mu}/\rho_{e\mu}$ or $\eta'_{e\mu}/(\rho'_{e\mu}-1)$.
Indeed, these ratios can be expressed as a function of the \New{appearance} amplitudes:
\begin{eqnarray}\label{eq:ratio-eta-rho}
	\frac{\eta_{e\mu}}{\rho_{e\mu}} &=& \frac{ \mathcal{I}m\left(-\frac{U_{e1}U^*_{\mu 1}}{U_{e3}U^*_{\mu 3}} \right)}{\mathcal{R}e\left(-\frac{U_{e1}U^*_{\mu 1}}{U_{e3}U^*_{\mu 3}} \right)} = \frac{\mathcal{I}m\left(\frac{U_{e1}U^*_{\mu 1}U^*_{e3}U_{\mu 3}}{U_{e3}U^*_{\mu 3}U^*_{e3}U_{\mu 3}} \right)}{\mathcal{R}e\left(\frac{U_{e1}U^*_{\mu 1}U^*_{e3}U_{\mu 3}}{U_{e3}U^*_{\mu 3}U^*_{e3}U_{\mu 3}} \right)}\nonumber\\
	&=& \frac{\mathcal{I}m\left(U^*_{\mu 1}U_{e1}U_{\mu 3}U^*_{e3} \right)}{\mathcal{R}e\left(U^*_{\mu 1}U_{e1}U_{\mu 3}U^*_{e3} \right)} \Lorenzo{= \frac{\mathcal{I}_{\mu e, 1 3}}{\mathcal{R}_{\mu e, 1 3}}},
\end{eqnarray}
since $U_{e3}U^*_{\mu 3}U^*_{e3}U_{\mu 3} = \vert U_{e3}\vert ^2 \vert U_{\mu 3}\vert ^2$ is a real number.
One can recognize the ratio between the amplitude of the $\sin(2X_{13})$ term and the amplitude of the $\sin^2(X_{13})$ term \New{from equation \ref{eq:P_mu_e}} (up to a factor 2 and a minus sign since the amplitudes have opposite signs). \New{This} means that the linear relationship between $\eta_{e\mu}$ and $\rho_{e\mu}$ \New{is driven by} the $\nu_{\mu}\rightarrow \nu_e$ and $\bar{\nu}_{\mu}\rightarrow \bar{\nu}_e$ $X_{13}$ patterns.
Similarly, the ratio $\eta'_{e\mu}/(\rho'_{e\mu}-1)$ can be expressed as:
\begin{eqnarray}\label{eq:ratio-eta-rho-prime}
	\frac{\eta'_{e\mu}}{\rho'_{e\mu}-1} &=& \frac{ \mathcal{I}m\left(\frac{U_{e2}U^*_{\mu 2}}{U_{e3}U^*_{\mu 3}} \right)}{\mathcal{R}e\left(\frac{U_{e2}U^*_{\mu 2}}{U_{e3}U^*_{\mu 3}} \right)}\nonumber\\
	&=& \frac{\mathcal{I}m\left(U^*_{\mu 2}U_{e2}U_{\mu 3}U^*_{e3} \right)}{\mathcal{R}e\left(U^*_{\mu 2}U_{e2}U_{\mu 3}U^*_{e3} \right)} \Lorenzo{= \frac{\mathcal{I}_{\mu e, 2 3}}{\mathcal{R}_{\mu e, 2  3}}} ,
\end{eqnarray}
which can be constrained using only the measurements of $\mathcal{R}_{\mu e, 23}$ and $\mathcal{I}_{\mu e, 23}$. \New{The same geometrical constraints for the other five triangles are derived in the Appendix \ref{app:all-formulas}.}

\subsection{Application: standard and generic unitary PMNS matrix}
In the previous section we explained how the PMNS matrix elements can be combined to make geometrical unitarity constraints in the complex plane, in particular in the shape of circles and lines. But as we illustrated in Section \ref{sec:probability}, deviations from unitarity could only show up in the measurements if the parametrisation of $U$ is non-unitary. Here we will compare the usual parametrisation adopted by the experiments with a more versatile one from Ref. \cite{Ellis2020b}.

If we assume that $U$ is a $n\times n$ unitary matrix, it can be parametrised by $\frac{n\times(n -1)}{2}$ ($= 3$ for $n=3$) mixing angles, and $\frac{n \times (n + 1)}{2}$ ($= 6$ for $n=3$) phases. Three of these phases can be removed due to lepton field rephasing, and two of them (the Majorana phases) do not play any role in the oscillation. Thus, a unitary parametrisation of the PMNS matrix involves three mixing angles $\theta_{ij}$ and one physical phase $\delta$. The usual parametrisation adopted in the oscillation analyses is the following:
\begin{eqnarray} \label{eq:3by3unitarymatrixparam}
	U = \begin{pmatrix}
1& 0 & 0 \\
0 & c_{23} & s_{23} \\
0 & -s_{23}    & c_{23}
\end{pmatrix} \begin{pmatrix}
c_{13}& 0 & s_{13} e^{-i \delta} \\
0 & 1 & 0 \\
-s_{13} e^{i \delta} & 0   & c_{13}
\end{pmatrix} \begin{pmatrix}
c_{12}& s_{12} & 0 \\
-s_{12} & c_{12} & 0 \\
0 & 0   & 1 \end{pmatrix} ,
\end{eqnarray}
\noindent where $c_{ij} = \cos(\theta_{ij})$ and $s_{ij} = \sin(\theta_{ij})$. This parametrisation is unadapted to perform unitarity tests, because it will always verify the unitarity conditions defined by Eq. \eqref{eq:unit-norm-row}, 
\eqref{eq:unit-norm-column}, \eqref{eq:unit-closure-row}, and \eqref{eq:unit-closure-column} by construction.

\comments{A possible approach consists in considering the general lepton mixing matrix without unitarity.} \Lorenzo{Indeed, if we don't impose any constraint on $U$, there are in total 18 independent parameters (coming from the 9 complex matrix elements). Identically to the unitary case, five phases can be removed, which leaves us with 13 independent parameters for a general non-unitary parametrisation of $U$.}
We adopt the parametrization provided in Ref. \cite{Ellis2020b}: 
\begin{eqnarray}\label{eq:LMM-parametrization}
	U \equiv\left(\begin{array}{ll}
		\left|U_{e 1}\right|\left|U_{e 2}\right| e^{i \phi_{e 2}} & \left|U_{e 3}\right| e^{i \phi_{e 3}} \\
		\left|U_{\mu 1}\right|\left|U_{\mu 2}\right| & \left|U_{\mu 3}\right| \\
		\left|U_{\tau 1}\right|\left|U_{\tau 2}\right| e^{i \phi_{\tau 2}} & \left|U_{\tau 3}\right| e^{i \phi_{\tau 3}}
		\end{array}\right),
\end{eqnarray}
with the values of the 4 \New{CP-violating} phases $\phi_{e2}$, $\phi_{e3}$, $\phi_{\tau 2}$ and $\phi_{\tau 3}$ \comments{extracted from \Lorenzo{that} paper to estimate the values of the relevant oscillation amplitudes.}

\Lorenzo{Let us highlight the link between the phases from the parametrisations and the position of the apex of the unitarity triangle in the complex plane. The slope of the sides of the triangles are defined by Eq. \eqref{eq:ratio-eta-rho} and \eqref{eq:ratio-eta-rho-prime}. Translating this ratio in terms of the parametrisation in Eq. \eqref{eq:LMM-parametrization} gives:} 
\begin{eqnarray}
	\frac{\eta_{e\mu}}{\rho_{e\mu}} &\equiv& \frac{\mathcal{I}m\left(\vert U_{e1}\vert \vert U_{\mu 1}\vert \vert U_{e3}\vert e^{-i\phi_{e3}} \vert U_{\mu 3}\vert \right)}{\mathcal{R}e\left(\vert U_{e1}\vert \vert U_{\mu 1}\vert \vert U_{e3}\vert e^{-i\phi_{e3}} \vert U_{\mu 3}\vert \right)}\nonumber\\
	&=& \frac{\sin(-\phi_{e3})}{\cos (-\phi_{e3})} = - \tan (\phi_{e3}),\\
	\frac{\eta'_{e\mu}}{\rho'_{e\mu}-1} &\equiv& \frac{\mathcal{I}m\left(\vert U_{e2}\vert e^{i\phi_{e2}}\vert \vert U_{\mu 2}\vert \vert U_{e3}\vert e^{-i\phi_{e3}} \vert U_{\mu 3}\vert \right)}{\mathcal{R}e\left(\vert U_{e2}\vert e^{i\phi_{e2}} \vert U_{\mu 2}\vert \vert U_{e3}\vert e^{-i\phi_{e3}} \vert U_{\mu 3}\vert \right)}\nonumber\\
	&=& \frac{\sin(\phi_{e2}-\phi_{e3})}{\cos (\phi_{e2}-\phi_{e3})} =  \tan (\phi_{e2}-\phi_{e3}),
\end{eqnarray}
where the matrix element norms completely vanish from the equation. \Lorenzo{The same computation in terms of the usual unitary parametrisation from Eq. \eqref{eq:3by3unitarymatrixparam} gives:}

\begin{eqnarray}
	\frac{\eta_{e\mu}}{\rho_{e\mu}}
	&=& \frac{\sin(\delta)}{\cos(\delta) + \frac{s_{12} s_{13} s_{23}}{c_{12} c_{23}}},\\
	\frac{\eta'_{e\mu}}{\rho'_{e\mu}-1}
	&=& \frac{\sin(\delta)}{\cos(\delta) - \frac{c_{12} s_{13} s_{23}}{s_{12} c_{23}}},
\end{eqnarray}
which is in agreement with Ref.~\cite{Ellis2020}. These equations shows that in both cases the constraint on the slope of the apex is mainly driven by the CP-violating phases of the parametrisation. In other words, the amount of CP violation is governed by the slope, which is the ratio between the imaginary and real parts of the apex coordinates.

\section{\label{sec:implementation-stan} {Illustration} of the constraints on a unitarity triangle}

\New{In section \ref{sec:geometry} we illustrated the mathematical connection between geometrical constraints in the unitarity triangle and neutrino oscillation amplitudes. In this section we intend to verify this connection numerically, using the Bayesian inference software Stan \cite{Carpenter2017}. Since the constraints of the oscillation amplitudes are not usually provided by the collaborations, we will do a more qualitative study, showing what could potentially be done if all the neutrino oscillation amplitudes were to be measured independently.} \New{For the sake of the argument}, we will assume experimental constraints on these quantities based on the global fits values \cite{Ellis2020,Ellis2020b}.
Indeed, the authors conveniently provide numerical values for their $3\,\sigma$ constraints on the squared matrix elements $\vert U_{\alpha i}\vert ^2$; we will take the best fit values for computing the true values of the oscillation amplitudes.
\comments{Tables \ref{tab:mean-unitarity} gives the numerical values of the assumed truth for the oscillation amplitudes of interest.
\begin{table*}[htp]
	\centering
	\begin{tabular}{|c|c|c|c|c|c|}\hline
		\multicolumn{6}{|c|}{Disappearance constraints} \\ \hline \hline
		${A}_{e,12}$	& ${A}_{e,13}$	& ${A}_{e,23}$	& ${A}_{\mu,12}$	& ${A}_{\mu,13}$	& ${A}_{\mu,23}$	\\ \hline
		0.203476				& 0.014872				& 0.006622				& 0.0368095					& 0.045864 						& 0.203868	\\ \hline
	\end{tabular}
	\vspace*{0.5 cm}\\
	\centering
	\begin{tabular}{|c|c|c|c|}\hline
		\multicolumn{4}{|c|}{Appearance constraints} \\ \hline \hline
		$\mathcal{R}_{\mu e,13}$	& $\mathcal{I}_{\mu e,13}$	& $\mathcal{R}_{\mu e,23}$	& $\mathcal{I}_{\mu e,23}$	\\ \hline
		0.023270				& 0.0118569				& -0.034740				& -0.0119622						\\ \hline
	\end{tabular}
	\caption{\label{tab:mean-unitarity}Mean values of the amplitudes assumed for the constraints in the unitarity plane.
	\New{Left}: values used for the disappearance constraints (Fig. \ref{fig:pmns_circle}).
	\New{Right}: values used for the appearance constraints (Fig. \ref{fig:pmns_slope}).}
\end{table*}}

In addition, we assume a fixed $5~\%$ error on each amplitude measurement.
The decision to use arbitrary uncertainties instead of current ones arises from the desire to better observe their effects, particularly the intersection points between all the constraints. Let us point out that this precision is not completely unrealistic \New{for some of the oscillation amplitudes. For instance, the JUNO experiment recently published its first oscillation analysis after 59 days of data taking, with a total of 2379 inverse beta decay candidate events \cite{abusleme2025measurementreactorneutrinooscillations}. As we will explicit in section \ref{sect:experimental_strategy}, JUNO is directly sensitive to the oscillation amplitude $A_{e,12}$, and therefore the purely statistical error on this amplitude would currently be at around 2\%.}
\\
Clearly, the observations presented here should be further studied with more accurate estimations of the current and future precision on the parameters, but \Lorenzo{in this work} they \Lorenzo{are meant to} provide some insights.

\subsection{Implementation framework}

To confirm these observations, we use the Bayesian inference software called Stan \cite{Carpenter2017} to build an a posteriori distribution on the $\rho_{e\mu}$ and $\eta_{e\mu}$ variables.
\comments{Stan uses Hamiltonian Markov-chain Monte-Carlo techniques to efficiently explore high-dimension parameter spaces, which is useful in the case of complex geometrical shapes. }


\New{As discussed in Section IIIB, contrary to most authors, we don't perform our own global analysis of neutrino spectra. Instead, for the sake of the exercise, we assume here that the real and imaginary parts of PMNS matrix complex elements are free (meaning a flat prior) and only constrained by the oscillation amplitudes measurements, which are defined by these matrix elements.} \comments{The a posteriori distribution on the PMNS matrix elements therefore consists in a set of flat priors on these matrix elements and a global likelihood which is the product of normal distributions with the difference between the derived oscillation amplitudes and the experimental constraints as mean value and the relative error of $5~\%$.}
Thanks to the Stan framework, the constraints on the matrix elements can then be recasted into constraints on $\rho_{e\mu}+i\eta_{e\mu}$ and $\rho'_{e\mu}+i\eta'_{e\mu}$ as complex numbers.

\subsection{Disappearance amplitudes constraints}

\begin{figure}[ht]
	\centering
	\includegraphics[width=0.45\textwidth]{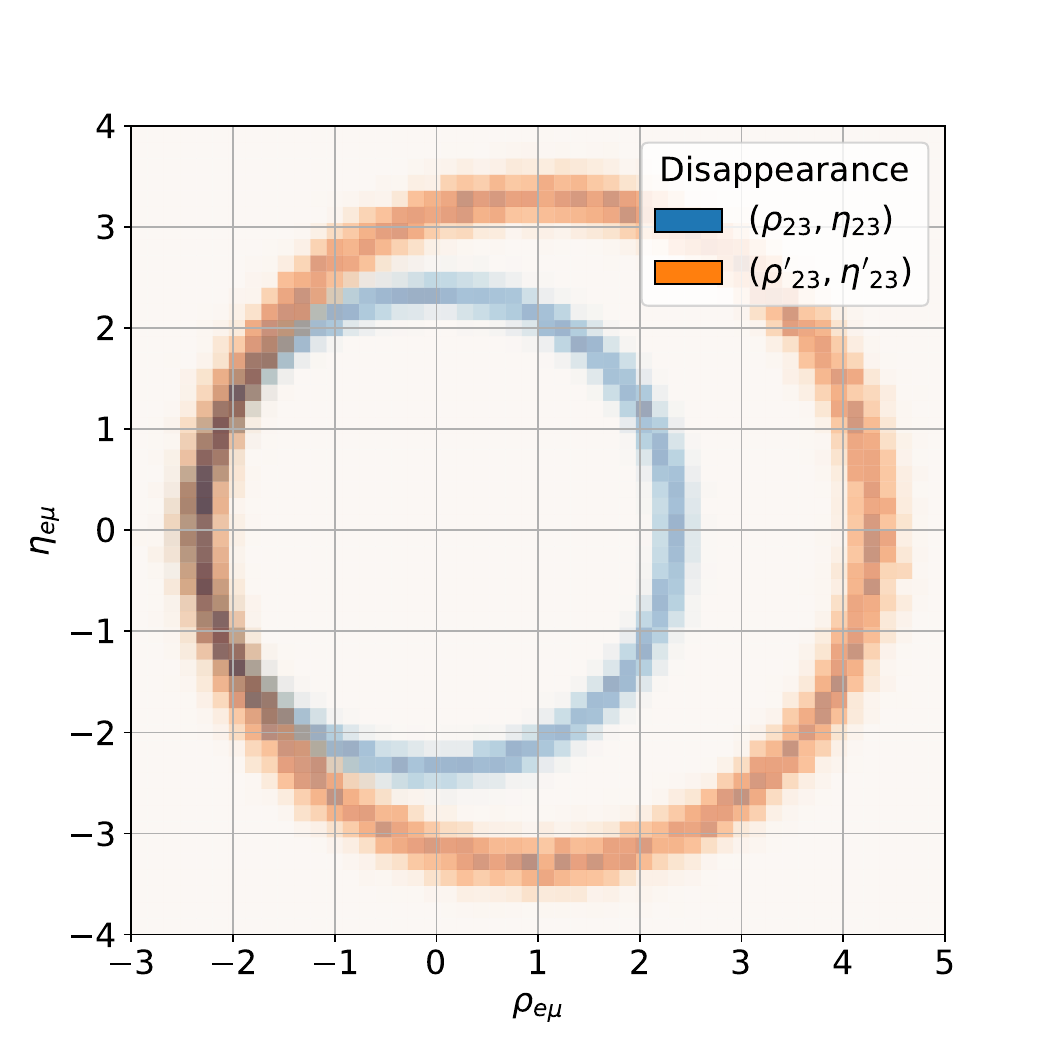}
	\includegraphics[width=0.45\textwidth]{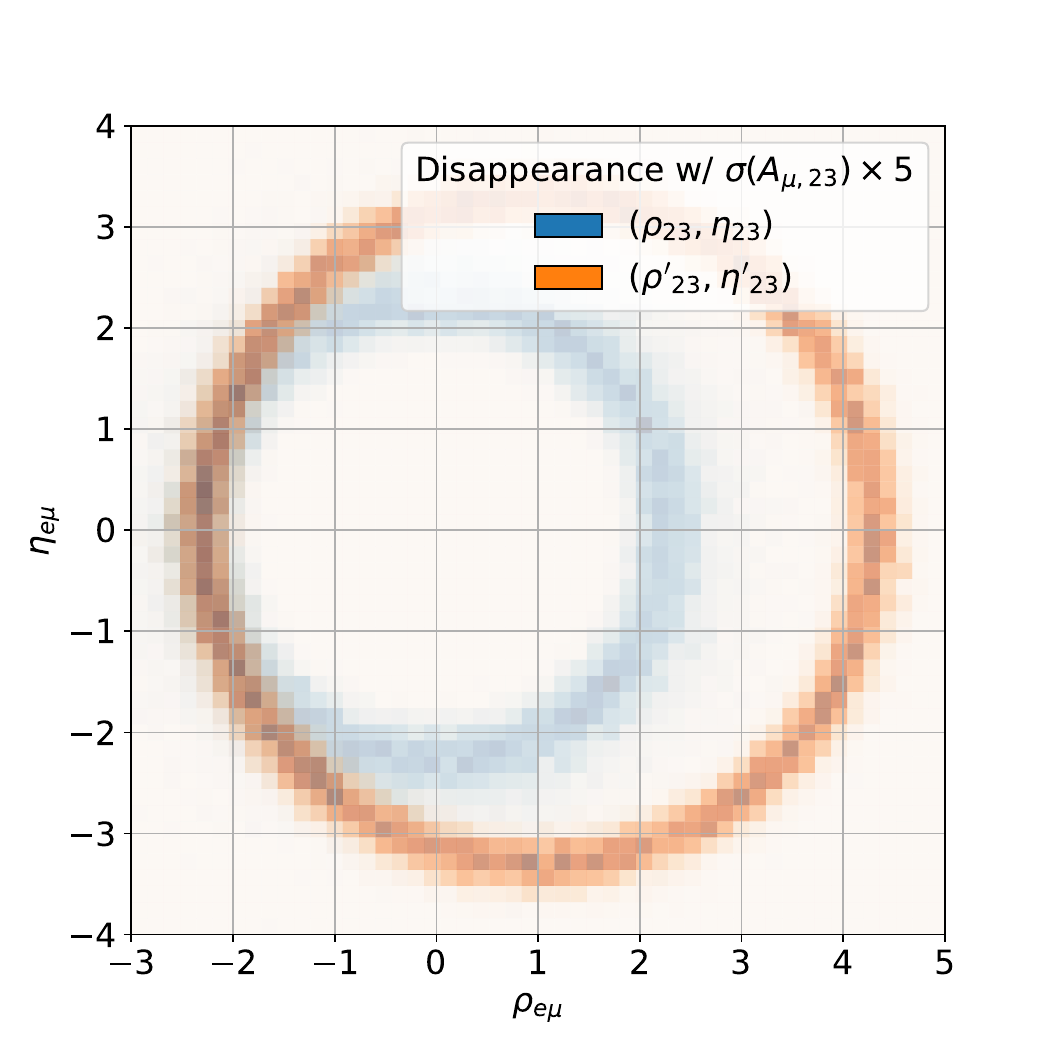}
	\caption{\label{fig:pmns_circle}Constraints on $(\rho_{e\mu},\eta_{e\mu})$ Eq.\eqref{eq:rho_eta_def} (blue) and $(\rho'_{e\mu},\eta'_{e\mu})$ Eq. \eqref{eq:rho_eta_prime_def} (orange) when considering only constraints on the disappearance amplitude measurements.
	\New{Left}: the amplitude values are computed using Table \ref{tab:mean-unitarity} and the error on each amplitude is set to $5~\%$.
	\New{Right}: Same, but where the error on $A_{\mu,23}$ is artificially increased by a factor of 5.}
\end{figure}

Figure \ref{fig:pmns_circle} shows the constraints with only $\nu_{\mu}$ and $\nu_e$ disappearance measurements e.g. $A_{\alpha,12}$, $A_{\alpha,13}$ and $A_{\alpha,23}$ with $\alpha\in\left\{e,\mu\right\}$.
As expected from Eq. \eqref{eq:unit_radius} and \eqref{eq:unit_radius_prime}, the constraint on each doublet $(\rho_{e\mu},\eta_{e\mu})$ and $(\rho'_{e\mu},\eta'_{e\mu})$ has a circular shape that only relates to the measurements of the 4 disappearance amplitudes.
The bottom plot of Fig. \ref{fig:pmns_circle} also shows the effect of the error on $A_{\mu,23}$: the radius uncertainty increases for the $(\rho_{e\mu},\eta_{e\mu})$ circular shape, while leaving $(\rho'_{e\mu},\eta'_{e\mu})$ unchanged \Lorenzo{as we expected from Eq.~\eqref{eq:unit_radius} and \eqref{eq:unit_radius_prime}}.
The measurement and combination of the 6 amplitudes for the $\nu_{\mu}$ and $\nu_e$ disappearance provide some information about the CP violation, as the intersection points between the blue and orange circles have a close to zero y-coordinate (although with an undetermined sign).
\New{These first results confirm the validity of Eq. \eqref{eq:unit_radius} and \eqref{eq:unit_radius_prime} and the fact that no other oscillation amplitude is necessary to obtain a circular constraint: in particular, no tau oscillation measurements are needed for this unitarity triangle.}
Additionally, contrary to Ref. \cite{Ellis2020,Ellis2020b}, the direct usage of the oscillation amplitudes instead of the mixing angles doesn't present the typical double-circle induced by the $\theta_{23}$ octant degeneracy, \New{which further strengthens the need for independent oscillation amplitudes measurements without using the reported mixing angle constraints.}

\subsection{Appearance amplitudes constraints}

Similarly, we can look at the impact of the appearance amplitudes constraints to the $(\rho_{e\mu},\eta_{e\mu})$ plane.

\begin{figure}[t]
	\centering
	\includegraphics[width=0.45\textwidth]{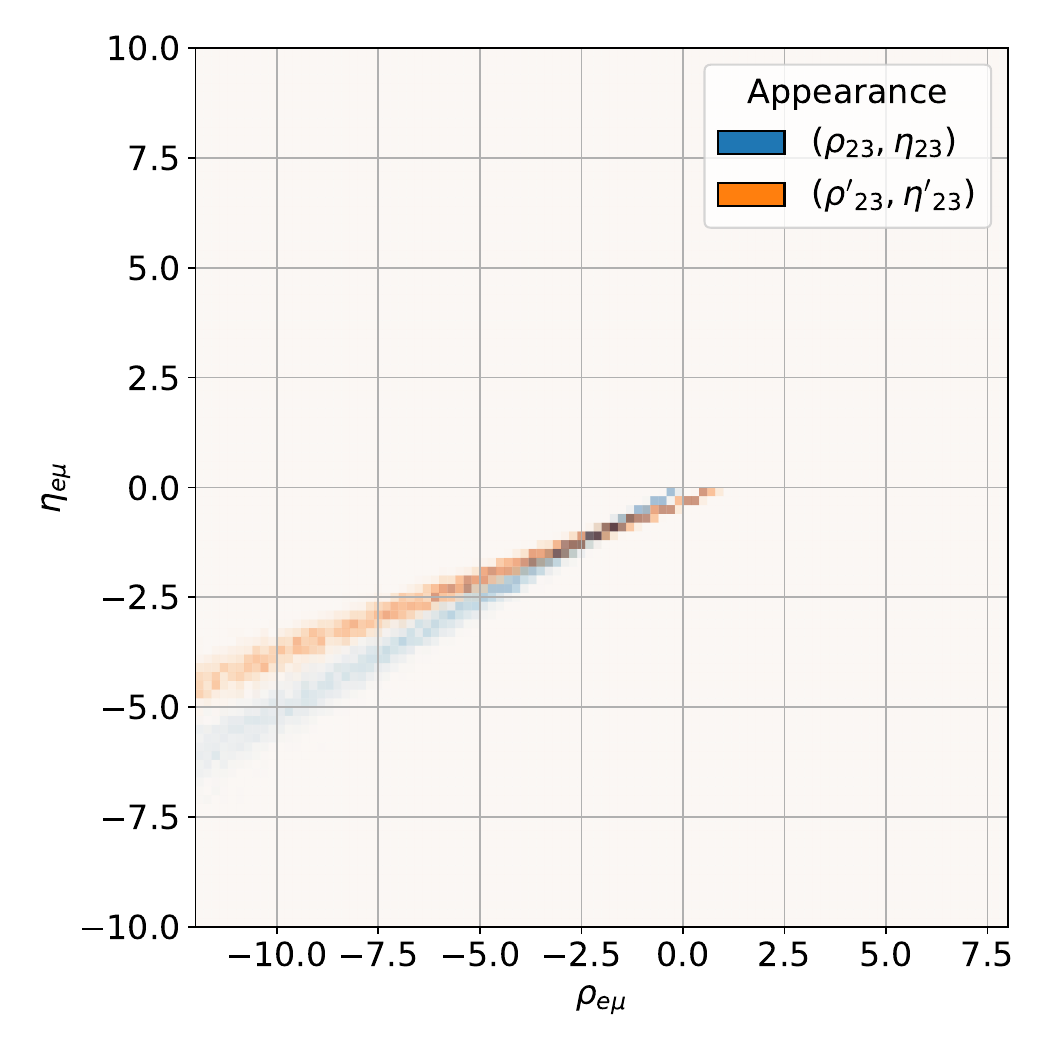}
	\includegraphics[width=0.45\textwidth]{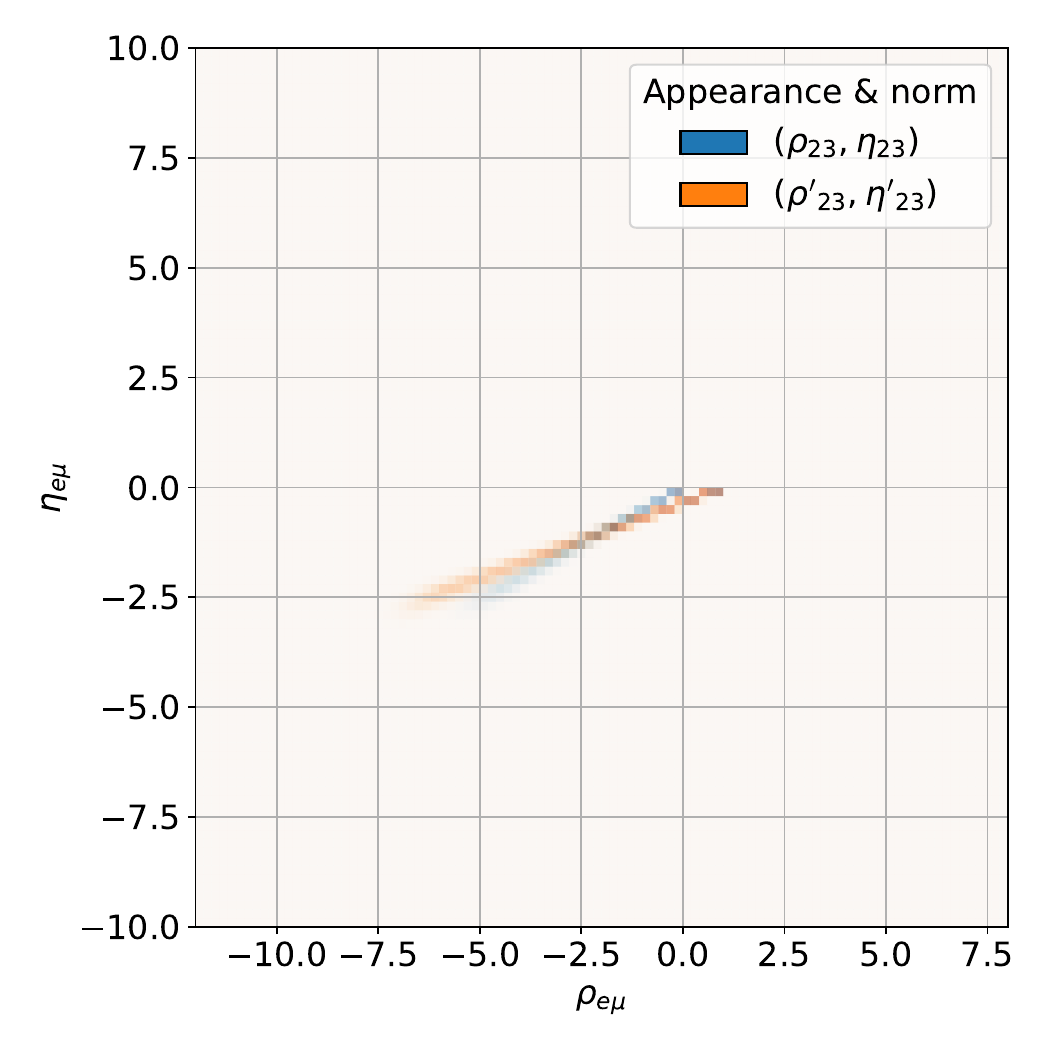}
	\caption{\label{fig:pmns_slope}Constraints on $(\rho_{e\mu},\eta_{e\mu})$ Eq. \eqref{eq:rho_eta_def} (blue) and $(\rho'_{e\mu},\eta'_{e\mu})$ Eq. \eqref{eq:rho_eta_prime_def} (orange) when considering constraints on the appearance amplitude measurements.
	\New{Left}: the amplitudes values are computed using Table \ref{tab:mean-unitarity} and the error on each amplitude is set to $5~\%$.
	\New{Right}: Same, but additional constraints on the $\left(\sum_{k}\vert U_{e k}\vert^2\right)^2$ and $\left(\sum_{k}\vert U_{\mu k}\vert^2\right)^2$ values are included as normal of mean value 1 and $5~\%$ error.}
\end{figure}

Table \ref{tab:mean-unitarity} gives the mean values of these oscillation amplitudes.
Assuming a typical error on these parameters of $5~\%$, it is possible to compute the expected range of the oscillation amplitudes and use them as constraints on the $(\rho_{e\mu},\eta_{e\mu})$ coefficients.
Figure \ref{fig:pmns_slope} \New{(left)} shows the expected constraints from the appearance amplitude measurements: as expected, each set of measurements $\mathcal{R}_{\mu e, 13}$ and $\mathcal{I}_{\mu e, 13}$ (resp. $\mathcal{R}_{\mu e, 23}$  and $\mathcal{I}_{\mu e, 23}$) provides radial-shaped constraints on $(\rho_{e\mu},\eta_{e\mu})$ (resp. $(\rho'_{e\mu},\eta'_{e\mu})$).
Since no constraint on the product of the matrix elements themselves is implied by these measurements, the line stretches from $(0,0)$ (or $(1,0)$) to infinity.

The additional constraints on e.g. the disappearance normalization values $\left(\sum_{k}\vert U_{e k}\vert^2\right)^2$ and $\left(\sum_{k}\vert U_{\mu k}\vert^2\right)^2$ allow to reduce the range of possible values of $\rho_{e\mu}$ and $\eta_{e\mu}$, as depicted in Fig. \ref{fig:pmns_slope} \New{(right)}.
This second set of constraints is more similar to the ones presented in \cite{Ellis2020b} as disappearance channel constraints, which indicates that the methodology employed there (a.k.a. the use of mixing angle constraints instead of oscillation amplitudes) implicitly includes normalization constraints from the appearance channels.

\subsection{Combined constraints}

\begin{figure}[t]
	\centering
	\includegraphics[width=0.49\textwidth]{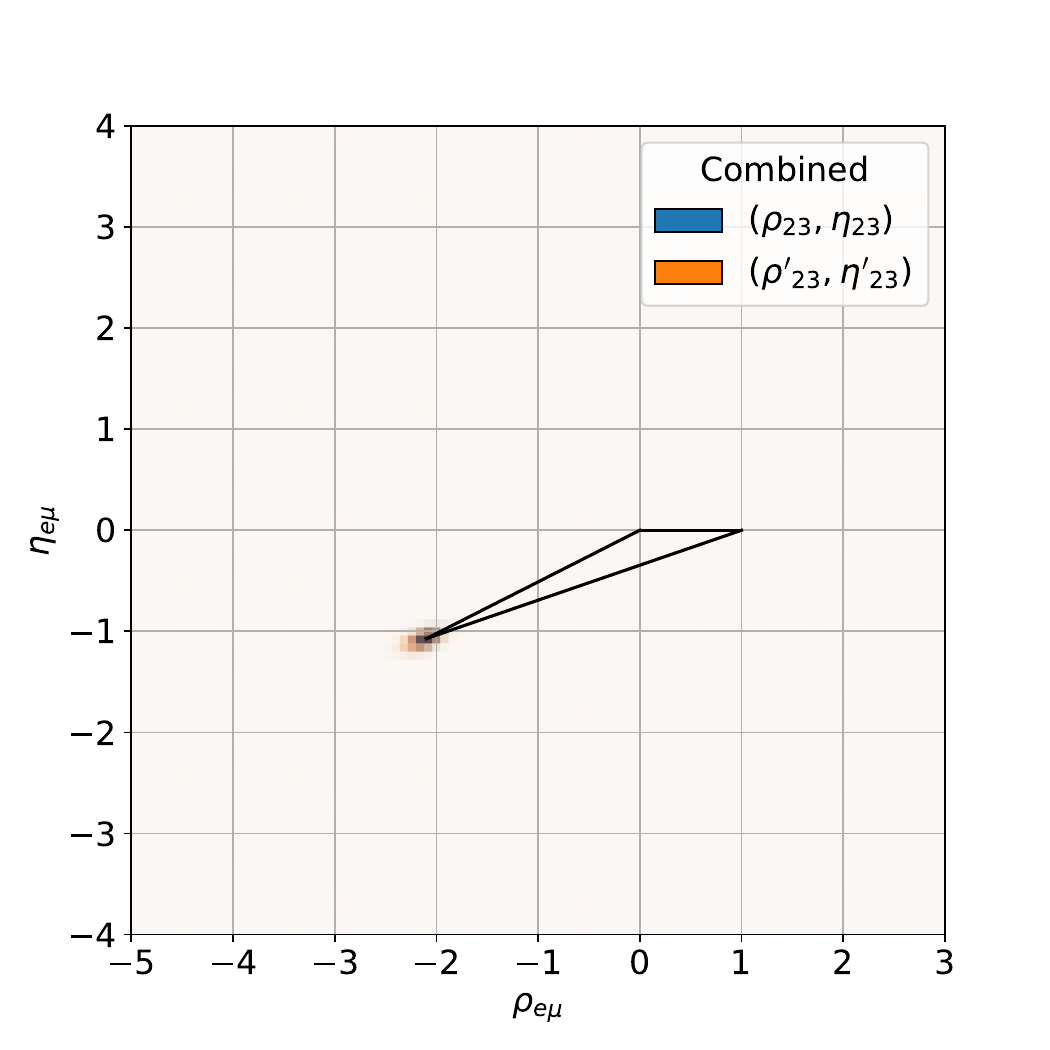}
	\includegraphics[width=0.49\textwidth]{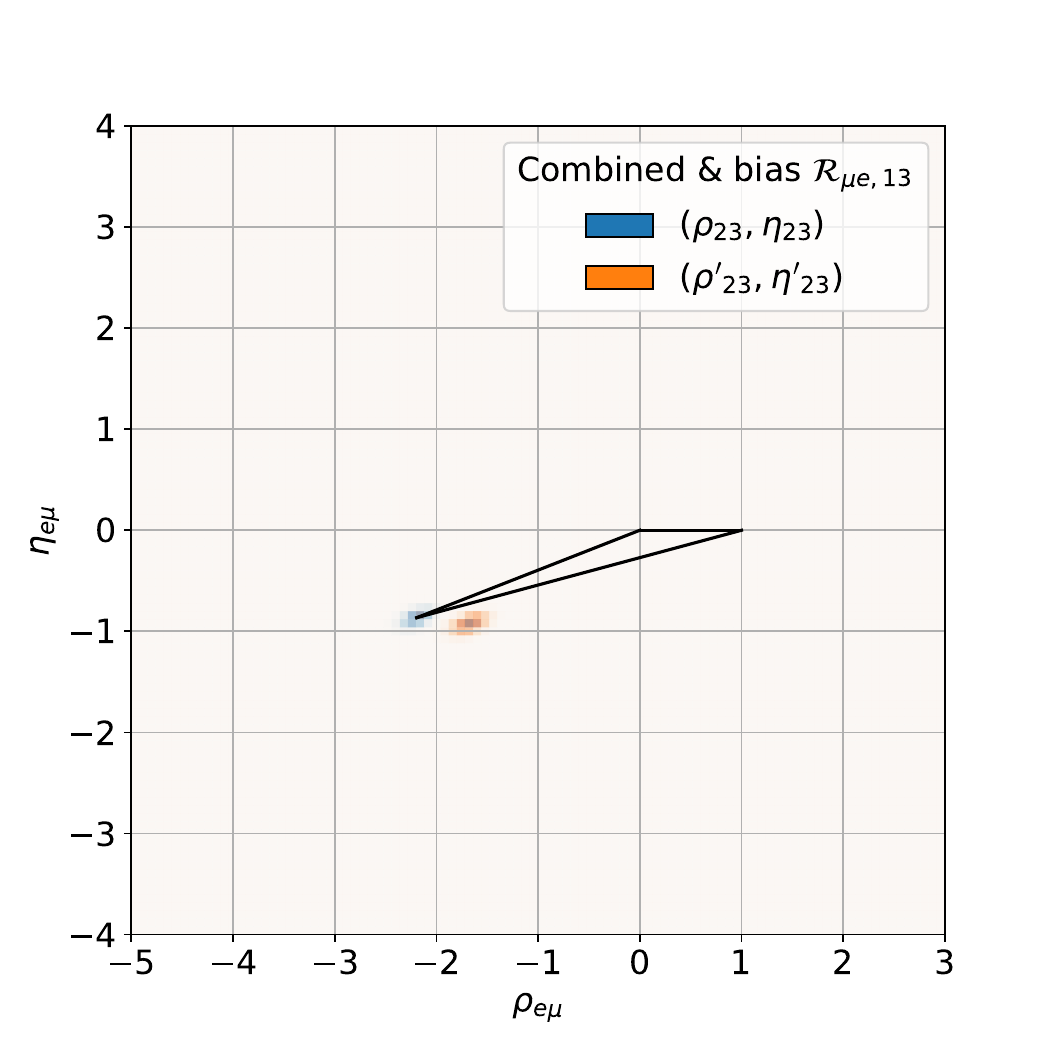}
	\caption{\label{fig:pmns_all}Constraints on $(\rho_{e\mu},\eta_{e\mu})$ Eq. \eqref{eq:rho_eta_def} (blue) and $(\rho'_{e\mu},\eta'_{e\mu})$ Eq. \eqref{eq:rho_eta_prime_def} (orange) when considering all constraints on the appearance and disappearance amplitudes measurements.
	\New{Left}: the amplitudes values are computed using Table \ref{tab:mean-unitarity} and the error on each amplitude is set to $5~\%$.
	\New{Right}: Same, but a bias of 50~\% is applied to $\mathcal{R}_{\mu e,13}$.}
\end{figure}

Figure \ref{fig:pmns_all} shows unitarity constraint when putting together the appearance and disappearance measurements.
All the constraints combine into a point-like shape for both $(\rho_{e\mu},\eta_{e\mu})$ and $(\rho'_{e\mu},\eta'_{e\mu})$ with perfect overlap; with worse constraints on the measurements, one expects a wider shape.
Figure \ref{fig:pmns_all} also shows the effect of a $50~\%$ bias on the disappearance amplitude $\mathcal{R}_{\mu e,13}$: in this case, the apexes do not overlap meaning a break of unitarity.

\section{\label{sec:discussion}Discussion}
\subsection{Practical limitations}

Let us \New{point out with} the observation that searching for a non-unitarity of the mixing matrix is way more complex in the context of neutrinos than for quarks.
Indeed, the quark sector possesses multiple probes of the CKM matrix elements and its various phases. 
For instance, it is possible to measure precisely the decay of mesons, corresponding to measuring the transition from one quark to another and so the absolute value of a single CKM matrix element.
Whereas the mostly accessible probe in the neutrino sector is their oscillation which involves between 2 and 4 PMNS matrix elements, \New{depending whether one considers the disappearance or appearance amplitudes}.

In practice, the experimental constraints on the oscillation amplitudes cannot be as clean as we assumed here; for example, the degeneracy between the $2-3$ and $1-3$ patterns in Eq. \eqref{eq:P_mu_mu}, \eqref{eq:P_e_e} and \eqref{eq:P_mu_e} induced by the close values of $\Delta m^2_{23}$ and $\Delta m^2_{13}$ means that one can experimentally measure primarily $\left( \vert U_{\mu 1}\vert^2+\vert U_{\mu 2}\vert^2\right) \vert U_{\mu 3}\vert^2$ from the oscillation pattern defined by Eq.~\eqref{eq:P_mu_mu}.
More realistic \New{sensitivity studies} should be \New{carried out} to better understand the constraints that one can have on the leptonic unitarity.

\Lorenzo{The studies performed in this work were made regarding the $(e, \mu)$ unitarity triangle. The primary reason is because the third row of the mixing matrix is hard to measure directly, as it involves $\tau$ neutrinos. Thus, for the moment, the primary focus should be to precisely measure independently the combination of the matrix elements of the first two rows of the mixing matrix, \New{via the experimental measurement of oscillation amplitudes}.}

\subsection{Experimental strategy\label{sect:experimental_strategy}}

This paper shows that certain combinations of oscillation amplitudes can give rise to geometrical constraints on the unitarity triangle apex. In particular, the same oscillation amplitudes combinations that intervene in the disappearance probability amplitudes, control the circle-like constraints, and in particular the radius from Eqs. \eqref{eq:unit_radius} and \eqref{eq:unit_radius_prime}. Similarly, the oscillation amplitudes that intervene in the appearance probability, can be used to constrain the slope of the radial-shaped constraints. Up until now we assumed that these combinations could be measured by the oscillation experiments individually. But in a more practical scenario, is this experimentally feasible?

Neutrino oscillation experiments \cite{Esteban2024} have shown the following hierarchy between the mass splitting: $|\Delta m_{12}^2| \ll |\Delta m_{13}^2| \simeq |\Delta m_{23}^2|$, with a difference of almost two orders of magnitude. Depending on the value of $L/E$, different terms can dominate in the oscillation probabilities. As an example, let us consider the electron neutrino disappearance probability from Eq.~\eqref{eq:P_e_e}. In the regime where $X_{12} \sim \frac{\pi}{2}$, the $X_{13}$ and $X_{23}$ oscillations are very rapid with $L/E$, so they are averaged out in the probability. Therefore, the dominant term is proportional to $|U_{e1}|^2 |U_{e2}|^2$. This is the regime where some electron disappearance experiments operate, like KamLAND and JUNO. But on the other hand, there is currently no muon neutrino disappearance experiment that operates in that regime, and that could therefore constrain $|U_{\mu1}|^2 |U_{\mu2}|^2$ directly. This is due to the practical difficulty to do such an experiment. \New{Indeed, this would also require $X_{12} \sim \frac{\pi}{2}$ but with energies strictly higher than the muon mass. The result would be either very low cross sections or extremely long baselines.}


In the regime where $X_{13} \sim X_{23} \sim \frac{\pi}{2}$, the $X_{12}$ term is too small to give rise to oscillations, so the dominant term is proportional to $|U_{e3}|^2(|U_{e1}|^2 + |U_{e2}|^2)$. Some electron disappearance experiments such as Daya Bay operate in this regime. It could be possible to disentangle between these two contributions if the second order terms are also constrained. Indeed, these terms are proportional to $|U_{e3}|^2 |U_{ei}|^2$ with $i = 1$ or $2$ depending on the mass ordering. In the muon neutrino sector, long baseline experiments such as T2K and NO$\nu$A operate in that same regime. Consequently their muon disappearance measurements are sensitive to $|U_{\mu3}|^2(|U_{\mu1}|^2 + |U_{\mu2}|^2)$ at first order. In the coming years, long-baseline experiments such as Hyper-Kamiokande and DUNE will be able to tightly constrain this quantity \New{thanks to the very large statistics of their muon disappearance samples}, and maybe even have access to the second-order terms, which are proportional to $|U_{\mu3}|^2|U_{\mu i}|^2$. 

Long baseline experiments also measure electron neutrino and antineutrino appearance. The electron neutrino appearance is described by the probability in Eq. \eqref{eq:P_mu_e}. In the regime $X_{13} \sim X_{23} \sim \frac{\pi}{2}$ the probability is driven by at least 3 terms: one proportional to $\sin^2(X_{i3})$ (with $i = 1$ or $2$), another proportional to $\sin(2 X_{i3})$ and the last one proportional to $\sin(2 X_{12})$. The oscillation amplitudes are also entangled in this case, like for the muon disappearance measurements. Consequently, in order to constrain them, other inputs are necessary, like the electron and muon  disappearance amplitudes.

In summary, if experiments such as Daya Bay are able to measure second-order combinations like $|U_{e3}|^2|U_{ei}|^2$, then all of the electron disappearance amplitudes $A_{e, ij}$ could be constrained. If not, then the matrix element amplitudes $|U_{ei}|$ have to be constrained individually. For this, solar neutrino experiments can be proven useful, such as SNO and Borexino, as they are sensitive to the quantity $|U_{e2}|^2 (|U_{e1}|^2 + |U_{e2}|^2) + |U_{e3}|^4$ at first order. For the muon row, things are more complicated because the measurements come almost exclusively from long baseline experiments. Therefore, there is no experiment capable of constraining $|U_{\mu1}|^2 |U_{\mu2}|^2$, and there are no inputs equivalent to the solar neutrinos for the electron row. To constrain the last muon disappearance amplitude $A_{\mu,12}$ and the appearance amplitudes used in equations \eqref{eq:ratio-eta-rho} and \eqref{eq:ratio-eta-rho-prime}, more complete feasibility studies should be performed by long-baseline \New{experiments}. Most probably the electron row measurements need to be used as inputs for this.

\section{Conclusions}

\New{An ideal scenario for unitarity testing in the lepton sector, would be if the experiments start reporting their constraints on the oscillation amplitudes.} Adopting a pedagogical approach more than a pure quantitative one, this paper explores an ideal scenario where the oscillation amplitudes \Lorenzo{are} independently measured  with a precision of $5~\%$.
Here we have demonstrated how the oscillation amplitudes measurements involving muon and electron neutrinos should be combined to draw \New{the} unitarity triangle \New{defined by $(\rho_{e\mu},\eta_{e\mu})$} and constrain unitarity.
Each geometrical figure in the $(\rho_{e\mu},\eta_{e\mu})$ plane is derived as a product or ratio of pure disappearance or pure appearance oscillation probabilities components.
\New{General expressions for the other unitarity triangles are also derived.}

\New{An important point developed in this paper is the fact that not all experimental constraints are required to constrain a given unitarity triangle. 
For instance, the current literature usually includes measurements involving tau neutrinos for the global fits that are used to construct the unitarity triangles. But as shown in Section \ref{sec:geometry}, the $\rho_{e\mu}+i\eta_{e\mu}$ constraints only need electron and muon neutrino measurements.
}

\New{To construct unitarity triangles for neutrinos,  the oscillation experiments are encouraged to measure and report the amplitudes of multiple oscillations terms, in order to disentangle the contributions from each term.
Additionally, if instead one had access to the measurement of a different combination of matrix elements or better to the measurement of individual elements, it would help draw different constraints on the triangles themselves. }

\section*{Acknowledgments}

The authors would like to thank Boris Popov and Marco Martini for the helpful comments and the kind guidance.
This research was funded by IN2P3/CNRS and the French ``Agence nationale pour la recherche" under grant number ANR-21-CE31-0008.

\appendix

\section{\label{app:all-formulas}Circular and radial constraints for all unitarity triangles}

\New{
This section develops the mathematical relationship between the neutrino oscillation amplitudes and the circular and radial constraints for all 6 unitarity triangles.
We follow the same ``flavor-democratic'' convention as Ref. \cite{Ellis2020} for the unitarity triangles definition.
Table \ref{tab:all-triangles} summarizes these constraints.
}

\begin{table*}[]
    \centering
    \begin{tabular}{|c|c|c|c|c|}
    \hline
    Triangle definition  & $R^2$ & $\eta/\rho$  & $R'^2$ & $\eta'/(\rho'-1)$ \\
     \hline\hline
    $\rho^{(1)}_{e\mu}+i\eta^{(1)}_{e\mu} = -\frac{U_{e1}U^*_{\mu 1}}{U_{e3}U^*_{\mu 3}}$  & $\frac{A_{\mu,12}A_{e,12}}{A_{\mu,23}A_{e,23}}$  & $\frac{\mathcal{I}_{\mu e, 1 3}}{\mathcal{R}_{\mu e, 1 3}}$ & $\frac{A_{\mu,12}A_{e,12}}{A_{\mu,13}A_{e,13}}$ & $\frac{\mathcal{I}_{\mu e, 2 3}}{\mathcal{R}_{\mu e, 2 3}}$ \\
    $\rho^{(2)}_{e\tau}+i\eta^{(2)}_{e\tau} = -\frac{U_{e2}U^*_{\tau 2}}{U_{e1}U^*_{\tau 1}}$  & $\frac{A_{e,23}A_{\tau,23}}{A_{e,13}A_{\tau,13}}$  & $\frac{\mathcal{I}_{\tau e, 21}}{\mathcal{R}_{\tau e,21}}$ & $\frac{A_{e,23}A_{\tau,23}}{A_{e,12}A_{\tau,12}}$ & $\frac{\mathcal{I}_{\tau e, 31}}{\mathcal{R}_{\tau e,31}}$ \\
    $\rho^{(3)}_{\mu\tau}+i\eta^{(3)}_{\mu\tau} = -\frac{U_{\mu3}U^*_{\tau 3}}{U_{\mu2}U^*_{\tau 2}}$  & $\frac{A_{\mu,13}A_{\tau,13}}{A_{\mu,12}A_{\tau,12}}$  & $\frac{\mathcal{I}_{\mu\tau, 23}}{\mathcal{R}_{\mu\tau,23}}$ & $\frac{A_{\mu,13}A_{\tau,13}}{A_{\mu,23}A_{\tau,23}}$  & $\frac{\mathcal{I}_{\mu\tau, 23}}{\mathcal{R}_{\mu\tau,23}}$ \\
    $\rho^{(-1)}_{12}+i\eta^{(-1)}_{12} = -\frac{U_{e1}U^*_{e 2}}{U_{\mu1}U^*_{\mu 2}}$  & $\frac{A_{e,12}}{A_{\mu,12}}$  & $\frac{\mathcal{I}_{\mu e, 12}}{\mathcal{R}_{\mu e,12}}$ & $\frac{A_{e,12}}{A_{\tau,12}}$  & $\frac{\mathcal{I}_{\tau e, 12}}{\mathcal{R}_{\tau e,12}}$ \\
    $\rho^{(-2)}_{13}+i\eta^{(-2)}_{13} = -\frac{U_{\mu1}U^*_{\mu 3}}{U_{\tau1}U^*_{\tau3}}$  & $\frac{A_{\mu,13}}{A_{\tau,13}}$  & $\frac{\mathcal{I}_{\tau\mu , 13}}{\mathcal{R}_{\tau\mu ,13}}$ & $\frac{A_{\mu,13}}{A_{e,13}}$ & $\frac{\mathcal{I}_{ e \mu , 13}}{\mathcal{R}_{e \mu,13}}$ \\
    $\rho^{(-3)}_{23}+i\eta^{(-3)}_{23} = -\frac{U_{\tau2}U^*_{\tau 3}}{U_{e2}U^*_{e 3}}$  & $\frac{A_{\tau,23}}{A_{e,23}}$  & $\frac{\mathcal{I}_{ e\tau, 23}}{\mathcal{R}_{e\tau,23}}$ & $\frac{A_{\tau,23}}{A_{\mu,23}}$ &  $\frac{\mathcal{I}_{ \mu\tau, 23}}{\mathcal{R}_{\mu\tau,23}}$\\
         \hline
    \end{tabular}
    \caption{Mathematical relationship between oscillation amplitudes and geometrical constraints on the unitarity triangles.}
    \label{tab:all-triangles}
\end{table*}

\New{Two key observations can be made regarding these results.\\
First of all, the expression for the circular constraints for the column unitarity triangles $\rho_{ij}^{(k)} + i\eta_{ij}^{(k)}$ only involve 2 oscillation amplitudes whereas four are required for the row unitarity triangles.
This comes from the fact that the neutrino oscillation requires two different mass eigenstates which naturally appears when considering the columns of the mixing matrix.
}

\New{Secondly, it should be worth noting that the column triangles all involve tau neutrino oscillation amplitudes. 
Even if, in appearance, it should therefore be simpler to constrain column unitarity triangles as less measurements are required, it is experimentally challenging to measure the tau neutrinos disappearance oscillation amplitude $A_{\tau,ij}$.}

\bibliographystyle{lemieuxdumonde_numbers}
\bibliography{HDR,biblio}
\end{document}